\documentclass[onecolumn,superscriptaddress]{revtex4-2}
\usepackage{caption}
\usepackage{subcaption}
\usepackage{mathtools,relsize}
\usepackage[]{graphicx}
\usepackage{appendix,physics,wasysym,footnote}
\usepackage[margin=1.55in]{geometry}
\usepackage{setspace,enumitem} 
\usepackage[english]{babel}
\usepackage{color,xcolor,amstext,amsmath,amssymb,amsfonts,dsfont}
\usepackage{euscript,braket}
\usepackage{bm,bbm}
\usepackage{hyperref}
\hypersetup{colorlinks,linkcolor={red},citecolor={blue},urlcolor={blue}}

\newcommand{\bla}{\color{black}}

\definecolor{tealgreen}{rgb}{0.0, 0.51, 0.5}

\newtheorem{Theorem}{Theorem}
\newtheorem{lemma}{Lemma}

\newenvironment{proof}{{\bf Proof:}}{\hfill$\square$}

\usepackage{lmodern}

\begin{document}
	\title{Quantum walks under superposition of causal order} 
	\author{Prateek Chawla}  
	\email{prateek.chawla@aalto.fi}
	\affiliation{The Institute of Mathematical Sciences, C.I.T. Campus, Taramani, Chennai - 600113, India.}
	\affiliation{Department of Applied Physics, Aalto University, Finland.}
	\author{Shrikant Utagi} 
	\email{shrikant.phys@gmail.com} 
	\affiliation{The Institute of Mathematical Sciences, C.I.T. Campus, Taramani, Chennai - 600113, India.}
    \affiliation{Department of Physics, Indian Institute of Technology Madras, Chennai - 600036, India.}
	\author{C. M. Chandrashekar}
	\email{chandracm@iisc.ac.in}
	\affiliation{The Institute of Mathematical Sciences, C.I.T. Campus, Taramani, Chennai - 600113, India.}
	\affiliation{Homi Bhabha National Institute, Training School Complex, Anushakti Nagar, Mumbai 400094, India}
	\affiliation{Quantum Optics \& Quantum Information, Department of Instrumentation \& Applied Physics, Indian Institute of Science, Bengaluru - 560012, India.}

	\begin{abstract}
	We set the criteria under which superposition of causal order can be incorporated in to quantum walks. In particular, we show that only periodic quantum walks or those with at least one disorder exhibit Superposition of causal order under the action of `quantum switch'. We exemplify our results with a simple example of two-period discrete-time quantum walks. In particular, we observe that periodic quantum walks exhibit causal asymmetry pertaining to the dynamics of the reduced coin state: the dynamics are more non-Markovian for one temporal order than the other. We also note that the non-Markovianity of the reduced coin state due to indefiniteness in causal order tends to match the dynamics of a particular temporal order of the coin state. We substantiate our results with numerical simulations.
	\end{abstract}
	
	\maketitle
	
	In the recent years, quantum walks have gained considerable interest as efficient tools to model controlled quantum dynamics \cite{meyer1996quantum,venegas2012quantum,chandrashekar2012disorder,mohseni2008environment,mallick2017neutrino,chawla2019quantum,innocenti2017quantum}. Much like a classical random walk, quantum walks also admit discrete-time and continuous-time realizations. The discrete-time quantum walk (DTQW) is defined on a composite Hilbert space consisting of the two, `coin' and `position' Hilbert space where the evolution is defined using a repeated application of a quantum coin operation applied on the coin space, followed by the coin dependent position shift operation on the composite space \cite{meyer1996quantum}. The continuous-time quantum walk (CTQW), however, is defined solely on the position space, with the evolution operator being dependent on the  construction of the position space \cite{gerhardt2003continuous}. Both variants of quantum walks have been used to  develop various quantum algorithms to perform various tasks,  schemes for quantum simulation \cite{godoy1992quantum,kitagawa2010exploring,chandrashekar2013two,chandrashekar2011disordered,chawla2020discrete,farhi1998quantum,konno2005one,yin2008quantum,douglas2008classical,kollar2012asymptotic}, and for realizing  universal quantum computation \cite{childs2009universal,lovett2010universal,singh2021universal,chawla2020multi}. A  quadratically faster  spread in the probability distribution of the walker in position  space is also shown by both variants in comparison to their classical counterpart  \cite{aharonov1993quantum,childs2003exponential}. The quadratic spread is readily available for use as a resource for quantum algorithms, and demonstrates the viability of quantum walks in implementation of quantum strategies that show speedup over their classical counterparts. In this work, we restrict ourselves to discrete-time quantum walks.

     Our comprehension of physical phenomena generally assumes that events happen in fixed causal order. Interestingly, due to recent advances in quantum foundations, it has been shown that linear quantum mechanics allows for a superposition of temporal (or, causal) order of quantum operations and  between any two events in general. The investigation of this phenomenon has been a topic of extensive research in the areas of quantum information and physics lately, both theoretically \cite{brukner2014quantum,ebler2018enhanced} and experimentally \cite{procopio2015experimental,rubino2022experimental}. One of the major possibilities brought out by this phenomena is that of investigation and simulation of quantum phenomena in a spacetime without a definite causal structure, for example in regions where quantum gravity effects are prominent. Interestingly, the so-called quantum switch is known to be a superoperator capable of creating superposition of causal order \cite{ebler2018enhanced}, and has recently been experimentally realized in photonic settings \cite{procopio2015experimental,taddei2021computational,rubino2022experimental}.
     
    Quantum walks are interesting mathematical models useful for applications ranging from physics to computer science, and in this work, we illustrate that a controllable quantum dynamics akin to a quantum walk is also possible in spacetimes which allow for a superposition of causal orders, and therefore this generalized dynamics (herein called quantum walk under a superposition of causal orders), also exhibits ballistic propagation, and may be used for quantum algorithms as such. We stress here that quantum walks under a superposition of causal orders cannot be described without a spacetime that permits operations with superposed temporal orders, and thus are the generalization of quantum walks in such a spacetime.
	
	A walker executing a one-dimensional DTQW is characterized by a coin operation and a position shift operation in the Hilbert space $\mathcal{H}_w = \mathcal{H}_c \otimes \mathcal{H}_p$, where $\mathcal{H}_c$ and $\mathcal{H}_p$ are the coin and the position Hilbert spaces, respectively. The position space basis is chosen to be columns of the identity matrix, with the basis states given by $\big\{\ket{x}, \; x\in \mathbb{Z} \big\}$. The coin space is a finite-dimensional Hilbert space which, in this case, is chosen to be 2-dimensional. The basis of the coin space is chosen to be the set $\big\{ \ket{0}, \ket{1}\big\}$. With the quantum coin operation defined as $C(\theta) \in SU(2)$, the $t$-step evolution is defined as, 
	\begin{equation}
		\ket{\psi(t)} = \big[S \big( C(\theta) \otimes \mathds{1}_p\big) \big]^t \ket{\psi(0)},
		\label{eq:dtqwops}
	\end{equation}
	\begin{equation*}
		\begin{split}
			\text{where,} \quad  &C(\theta) = \begin{bmatrix}
				\cos(\theta) & i \sin(\theta) \\
				i \sin(\theta) & \cos(\theta)
			\end{bmatrix}, \\
			\text{and } S =& \sum_{x\in \mathbb{Z}} \bigg[ \ket{0}\bra{0} \otimes \ket{x-a}\bra{x}  + \ket{1}\bra{1} \otimes \ket{x+b}\bra{x}  \bigg].
		\end{split}
	\end{equation*}
	
	\noindent Here, $a,b \, \in\, \mathbb{Z}$ represent the amount of traversal in the position space experienced by the component of walker's probability amplitude in the eigenspaces corresponding to the coin space basis vectors $\ket{0}$ and $\ket{1}$, respectively. In this work, for the sake of clarity and convenience, we choose $a=b=1$. The initial state of the walker, $\ket{\psi(0)}$ 
	is typically chosen to be localized onto a single eigenstate of $\mathcal{H}_p$, and an equal superposition of eigenstates in $\mathcal{H}_c$, i.e. 
	\begin{equation}
		\ket{\psi(0)} = \frac{1}{\sqrt{2}}\left( \ket{0}+\ket{1} \right) \otimes \ket{x=0}. 
		\label{eq:initstate}
	\end{equation}
	\noindent
	Canonically, the coin rotation angle $\theta$ is assumed to lie between $[0,\pi]$ unless specified otherwise.

 Let $U_1$ and $U_2$ be two unitary operators, then the action of the quantum switch, denoted $\mathcal{S}$, is given by
	\begin{align}
		& \mathcal{S}(U_1, U_2)[\rho_s \otimes \rho] =  \mathcal{U}(\rho_s \otimes \rho )\mathcal{U}^\dagger
		\label{eq:indef-unitary}
	\end{align}
	where,
	\begin{align*}
		\mathcal{U} = \ket{0}\bra{0} \otimes U_1 U_2 + \ket{1}\bra{1} \otimes U_2 U_1 ,
	\end{align*}
	which acts on the joint Hilbert space of the `target state' $\rho$ and the switch state $\rho_s$. When the switch state is $\rho_s =\ket{+}\bra{+}$, the implementation of dynamics described by Eq.~(\ref{eq:indef-unitary}) results in $U_1$ and $U_2$ being applied to $\rho$ in Superposition of causal order, whereas when $\rho_s$ is in the state $\ket{0}\bra{0}$ or $\ket{1}\bra{1}$, the order of the operators $U_1 U_2$ or $U_2 U_1$ is implemented, respectively. It can be shown that this manifests as a task of identifying if two unitaries commute or not, and such a situation has lead to understanding the relationship between Superposition of causal order and quantum computational speed up.
	
	Superposition of causal order has been shown to be a resource for quantum computation and communication \cite{chiribella2021indefinite,abbott2020communication}, while in some cases it need not be so \cite{jia2019causal}. On the other hand, quantum non-Markovianity and causality are  other  topics that have attracted attention because of their intimate relationship in physical phenomena \cite{milz2018entanglement,utagi2021quantum,giarmatzi2021witnessing}. In the case of discrete-time quantum walks, the dynamics of the coin state can be interpreted as if it is passing through a quantum channel $\Phi$ induced by the quantum walk, and which can also be given a Kraus representation \cite{naikoo2020non}. \bla In the present work, we consider the effect of superposition of causal order in discrete-time quantum walks. We prove that a discrete-time quantum walk can exhibit nontrivial dynamics if it is either periodic or contains temporal disorder in coin operations. Here, nontrivial behaviour implies that the resulting dynamics cannot be written as another quantum walk of the same form as the original quantum walk under consideration. In a discrete-time quantum walk with two coin operations, the coin state can be interpreted as passing through channels $\Phi_1$ or $\Phi_2$, corresponding to its causal ordering, and Superposition of causal order in which case the effective channel is given by Eq.~(\ref{eq:indef-unitary}).  In each case, we see varied amounts of non-Markovian behavior in the reduced dynamics of the coin state $ \rho_c \in \mathcal{H}_c$, in the sense that we observe a `causal asymmetry' \cite{thompson2018causal} in the dynamics pertaining to the degree of non-Markovianity of the coin state.

Now, we move to our main results.

	A two-period DTQW is a variant of DTQW where the walker at each step alternates between two different coins. The $N$-step walk operation is then characterized {by} the following unitary:
	\begin{equation}
		\begin{split}
			U &=  (U_2)^{N \bmod 2 } \left(U_1 U_2\right)^{\lfloor N/2 \rfloor}, \\
			\text{where,} & {} \\
			U_1 &= SC_1, \quad \text{~~and~~} \quad U_2 = SC_2.
		\end{split}
		\label{eq:QW2pd}
	\end{equation}
	\noindent
	Here, $U_1$, $U_2$ denote a single walk step, $S$ is the shift operator as shown in Eq.~(\ref{eq:dtqwops}), and $C_1, C_2$ are single-parameter coin operators, where the subscript is used to differentiate the parameters, i.e. $ C_j \equiv C(\theta_j) \otimes \mathds{1}$. Here, $\lfloor N \rfloor$ denotes the greatest integer less than or equal to $N$. As is apparent from the form of Eq.~(\ref{eq:QW2pd}), the 2-period DTQW in reverse temporal order is equivalent to exchanging the two coin operations if $N$ is even. In fact, for two-period walks, there can be only two sequences (or, permutations) of coin operations which do not commute with each other, allowing for a superposition of causal order in quantum walk unitaries. {This can be seen from a more generalized point of view, as a DTQW is a unitary operator that can be expressed as an ordered sequence of one or more one-step DTQWs, also known as `DTQW steps'. 
		
		We have the following theorem. 
		
		\begin{Theorem}
			A quantum walk exhibits nontrivial dynamics under superposition of its forward and reverse causal orders iff at least two of the DTQW steps do not commute. \bla (Here, `trivial dynamics' implies that the effective dynamics may be written as a quantum walk of the same form as the original DTQW under consideration.) \bla
        \label{th:nontrivialdynamics}
		\end{Theorem}
		\begin{proof}
			We prove this theorem by contradiction as follows.
			Let an $N$-step quantum walk be denoted by the unitary $U := \Pi_{i=0}^{N-1}U_i $, where $U_i \equiv S[C(\theta_i)\otimes\mathds{1}] $, and $\theta_i \in \mathbb{R}$. We assume that $[U_j,U_k]=0$ $\forall j,k \in \big\{0,1,\cdots,N-1 \big\}$, and the walk $U$ exhibits nontrivial dynamics when executed in superposition of causal order. Since $[U_j,U_k]=0$ $\forall j,k \in \big\{0,1,\cdots,N-1 \big\}$, it implies that for $U' := \Pi_{i=0}^{N-1}U'_i$, where $U'_i = U_{N-i-1}$, $i \in \big\{0,1,\cdots,N-1 \big\}$, we have that $U = U'$. Therefore, the operation $\mathcal{U}$ as defined in Eq.~\eqref{eq:indef-unitary} reduces to the operation $U$ on the target, resulting in trivial dynamics. This contradicts the initial assumption, and proves the \textit{if} condition. The \textit{only if} statement may be proved in a similar manner. Since $[U_j,U_k]\neq 0$ for at least one pair $j,k \in \big\{0,1,\cdots,N-1 \big\}$, this implies that $U \neq U'$. Thus, the operator $\mathcal{U}$ cannot be reduced to a separable form, and the quantum walk will exhibit nontrivial dynamics.
		\end{proof}
		
		In order for quantum walks to exhibit Superposition of causal order, non-commutativity of walk unitaries is essential. Thus, we propose the following lemma:
		{\begin{lemma}
				Two DTQW steps having identical shift operations of the form shown in Eq.~(\ref{eq:dtqwops}) and different single-parameter coins characterized by the parameters $\theta_i$, where ($i=1,2$) shall only commute if $\theta_1$ and $\theta_2$ differ by an integral multiple of $\pi$.
				\label{lem:qwComm}
		\end{lemma}}
		\begin{proof}
			An elementary proof may be constructed by considering a matrix representation of the coin Hilbert space. The representation of the shift operation will then be,
			\begin{equation}
				\begin{split}
					S &= \sum_{x\in \mathbb{Z}} \big[ \ket{0}_c\bra{0} \otimes \ket{x-1}\bra{x} + \ket{1}_c\bra{1} \otimes \ket{x+1}\bra{x}  \big] \\
					&= \begin{bmatrix}
						\sum_{x\in \mathbb{Z}} \ket{x-1}\bra{x} & 0 \\
						0 & \sum_{x\in \mathbb{Z}} \ket{x+1}\bra{x}
					\end{bmatrix} \\
					&= \begin{bmatrix}
						T_- & 0 \\ 0 & T_+
					\end{bmatrix}
				\end{split}
			\end{equation} 
			where $T_\pm$ are the position space propagators of the walker. It is easily verified that $T_\pm = T_\mp^\dagger$ and $T_-T_+ = T_+T_- = \mathds{1}$ from the fact that $S$ is unitary. For this proof, we consider coin operator for the coin $C(\theta)$ as in Eq.~(\ref{eq:dtqwops}). Now, the commutator of the quantum walks will look like,
			{\small	\begin{equation}
					\begin{split}
						&{} \left[SC_1, SC_2\right] \\ 
						&= \begin{bmatrix}
							0 & i(\mathds{1}-T_+^2)\sin(\theta_2 - \theta_1) \\
							-i(\mathds{1}-T_-^2)\sin(\theta_2 - \theta_1) & 0
						\end{bmatrix}
					\end{split}
			\end{equation}}
			\noindent
			Thus, we obtain that,
			\begin{equation}
				\begin{split}
					&\left[SC_1, SC_2\right] = 0 \\
					\implies   &\sin(\theta_2 - \theta_1) = 0 \\
					\implies &\theta_2 = \theta_1 + n\pi, \qquad n \in \mathbb{Z},
				\end{split}
			\end{equation}
			which completes the proof.
		\end{proof}
		
		From the above, we may infer that  the only DTQWs capable of exhibiting Superposition of causal order are either DTQWs with disorders or periodic quantum walks.
		
    \begin{figure}[!h]
		\centering
	\includegraphics[width=9cm]{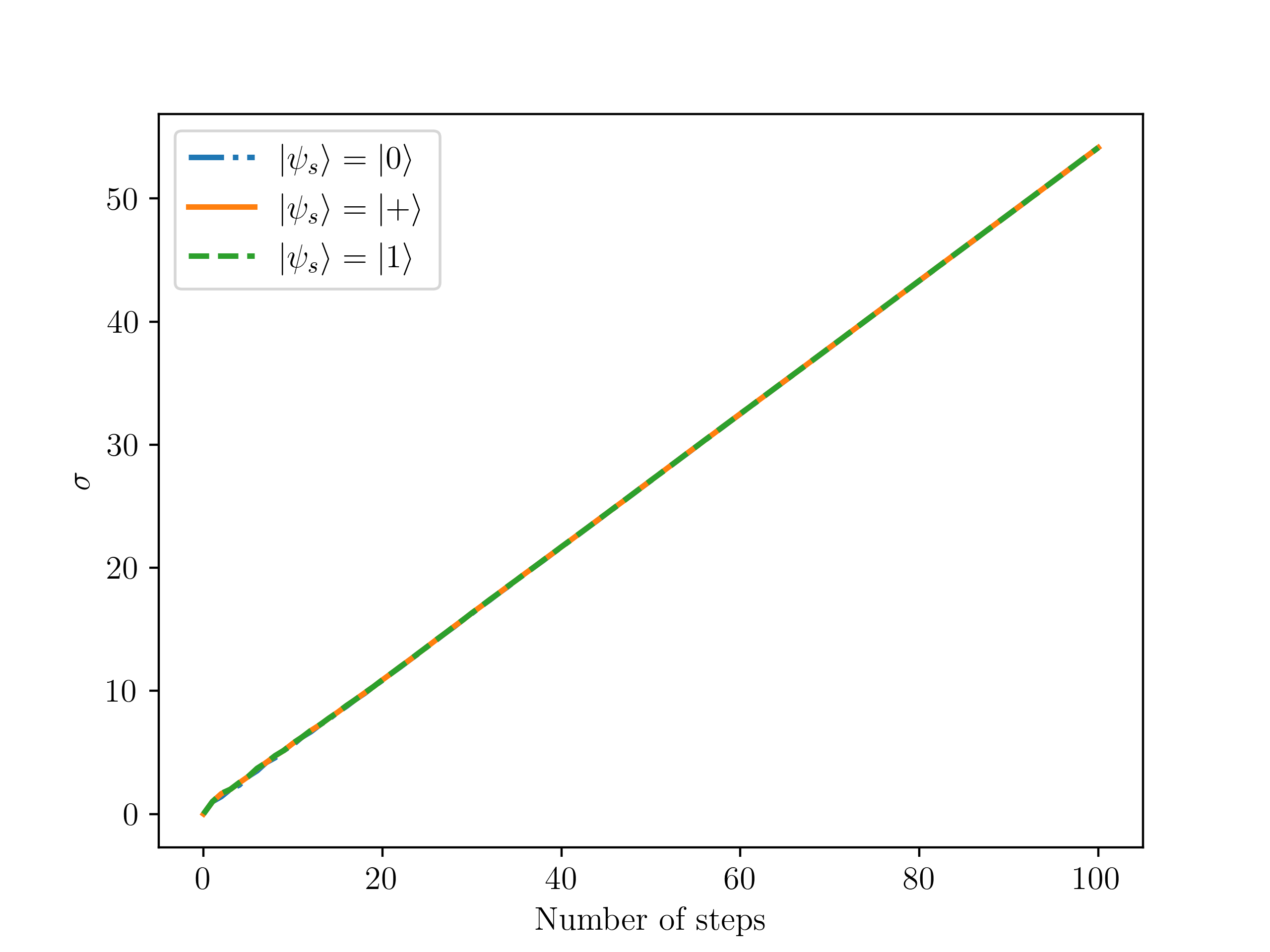}
		\caption{The plot of the spread of a quantum walker in definite (blue dashed line), reversed (green dashed line), and superposed (orange solid line) causal order of its steps. The walker executes a 2-period discrete-time quantum walk with a localized initial state $\left( \frac{\ket{0} + \ket{1}}{\sqrt{2}}  \right) \otimes \ket{x=0}$, for $100$ steps, as defined in Eq.~\eqref{eq:QW2pd}. The two coin angles $\left( \theta_1, \theta_2\right)$ are chosen to be $\frac{\pi}{6}$ and $\frac{\pi}{4}$, respectively. It is seen that the use of a quantum switch has no effect on the spreading of the quantum walker in its position space. This is true for multi-period DTQWs in general.}
		\label{fig:switchSpread}
    \end{figure} 
    
From the theorem \ref{th:nontrivialdynamics} above, it may be seen that for periodic quantum walks, one can implement Superposition of causal order in two distinct ways: (i) implementing each period in superposition of causal order, and (ii) implementing the entire sequence of steps of the quantum walk in superposition of temporal order of its steps.

It can be seen that the case (i) above corresponds to the repeated concatenation of a completely positive, trace preserving map on the initial state of the walker. The effective dynamics induced by this type of map are a form of temporal noise, and will simply cause the walker to localize \cite{chandrashekar2011disordered}. In this work, we consider only the case (ii) as defined above.

Here, we would like to clarify that we essentially consider a sequence of walk unitaries $U_f = U_1\cdot U_2\cdots U_N $, whose reverse temporal order is $U_r = U_N\cdot U_{N-1}\cdots U_1$. It is important to note here that executing the steps of a quantum walk in reversed temporal order is not equivalent to executing the quantum walk in reverse. The former implies that  quantum walk steps are executed on the initially localized state in reversed causal order, while the latter implies the application of the inverse operation of each step to the \textit{final} state of the walk at time $t=0$, to return to the initial state after $N$ steps. In other words, reversing the order of quantum walk steps results in a different periodic quantum walk. We dub this fact as `causal asymmetry' in periodic quantum walks. Note that neither the inherent causal asymmetry nor the execution of the quantum walk under Superposition of causal order has any effect on the spread of the walker's probability distribution in its position space. This can be seen in the Fig.~\ref{fig:switchSpread}. This can further be explained by the fact that for periodic quantum walks, the spread of the walker varies linearly with time, and is bounded by $\pm \min\left\{ N\cos(\theta_1), N\cos(\theta_2) \right\}$, where $\theta_i$ are the coin angles, and $N$ is the number of steps \cite{kumar2018bounds}. In our case, irrespective of the causal order (forward or reverse) in which the walk is executed, the two coin angles remain the same, and hence the spread of the walker is identical for both cases. It is, however, interesting to see that the ballistic spread of the reduced dynamics of the quantum walker in its position space is unaffected by superposition of its causal orders. 

This result is not trivial, and applies to periodic discrete-time quantum walks in general. Thus, superposition of a sequence of quantum walk steps with its own reverse is another class of controllable quantum dynamics which shows a ballistic spread in the position space of the walker. However, creating the superposition on the initial walker state requires an operation of Kraus rank 2, which cannot be expressed as dynamics resulting from a quantum walk, as the effective operation is no longer a unitary, but a completely positive, trace preserving map. For a mathematical analysis of why the Kraus rank increases due to superposition of causal orders, we refer the reader to Theorem \ref{thm:krausrankproducts} in Appendix \ref{ap:apA}. 

The causal asymmetry in dynamics becomes consequential when one considers the complementary case, \textit{i.e.}, the reduced dynamics of the walker in the coin space. It is known that quantum walks may be viewed as quantum channels for the coin qubit, and that these maps are non-Markovian \cite{naikoo2020non}. In the following subsection, we consider the effects of causal asymmetry and superposition of causal order on the non-Markovianity of the reduced dynamics of the coin.

	Quantum non-Markovianity is one of the topics that has attracted much attention of quantum information community. Various methods have been proposed to characterize and quantify quantum non-Markovianity \cite{rivas2014quantum,utagi2021quantum,li2018concepts}, yet its exact and complete treatment still remains an open problem. In this work, it suffices for us make use of the definition of non-Markovianity based on trace distance and measure the total memory in coin dynamics with a quantifier proposed by Breuer-Laine-Piilo \cite{breuer2009measure}.

	Here we give a brief account of how the reduced dynamics of the coin state can be computed in a simple way. In order to calculate the trace distance based measure of non-Markovianity, we consider two orthogonal initial coin states which are chosen to be $\ket{+}_c$ and $\ket{-}_c$. The initial density matrices of the (reduced) coin space are thus given by $\rho_+(0) = \ket{+}\bra{+}$ and $\rho_-(0) = \ket{-}\bra{-}$. The density matrices at any point $t$ are easily calculated by executing the walk for $t$ steps and tracing  out the position space from the resulting density matrix of the walker, i.e.
	\begin{align}
		\rho_c(t) = \mathrm{Tr}_{s,p}\left[ V \left(\rho_s \otimes \rho_c(0) \otimes\rho_p(0) \right) V^\dagger \right], 
		\label{eq:reducedcoinchannel}
	\end{align}
	\noindent
	where $\rho_p(0) = \ket{x=0}\bra{x=0}$, and $V$ is an operator, as given in Eq.~\eqref{eq:indef-unitary}, that represents the dynamics of quantum walks caused due to particular definite causal order and Superposition of causal order of coin operations.  \bla  In the present work, as noted in the previous sections, there can exist two cases: one, where each period of the periodic DTQW is put in superposition of causal order, and where the entire walk sequences are put in superposition of causal order. In this manuscript, however, we focus on the latter scenario. We see that for a $k-$period walk executed for $N$ steps, the sequence of the quantum walk steps for forward and reverse temporal orders is the same only when $N$ is  an integer multiple of $k$, and dissimilar otherwise. \bla
	
	The trace distance of two states $\rho_1, \rho_2$ can be calculated as
	\begin{align}
		D\left(\rho_1, \rho_2\right) = \frac{1}{2} \left\lVert \rho_1 - \rho_2 \right\rVert_1, 
		\label{eq:tracedistance}
	\end{align}
	where $\rho_1,\rho_2$ are a pair states in the same (or isomorphic) Hilbert space(s) and $\Vert A \Vert_1 = \Tr{\sqrt{A^\dagger A}}$ is the trace norm of an operator $A$. In other words, the states $\rho_{\pm}(0)$ described earlier may be interpreted as the initial states that pass through the coin channel (CPTP map) which is implicit in the Eq. (\ref{eq:reducedcoinchannel}), and  $\rho_\pm(t)$ are the states evolved through the quantum walk channel after $t$ discrete time steps.

	Now, the measure due to Breuer-Laine-Piilo (BLP) \cite{breuer2009measure} is simply an integral over the positive slope of the $D(t)$, which is given by
	\begin{align}
		\mathcal{N} = \max\limits_{\rho_1 , \rho_2} \int_{\frac{dD(t)}{dt} > 0 } \frac{dD(t)}{dt} dt, 
		\label{eq:BLPdef}
	\end{align}
	where $D(t) \equiv D\left(\rho_1(t), \rho_2(t)\right) $, as defined in Eq.~\eqref{eq:tracedistance}, and $\rho_{1,2}(t)$ are the two coin states at time $t$ [see Eq.~\eqref{eq:reducedcoinchannel}]. It is known that BLP measure requires optimization over the initial pair of states.  In our simulations, it was seen that this measure was maximized for the $\ket{+}$ and $\ket{-}$ states, hence the choice of our coin states. \bla At this point, it is important to note that BLP measure can give false positive values under a trace-decreasing CP map \cite{filippov2021trace}. Thus, we must re-normalize the dynamical map induced on the coin space at each step. \bla
	\begin{figure}
		\centering
		\begin{subfigure}{\linewidth}
			\includegraphics[width=9cm]{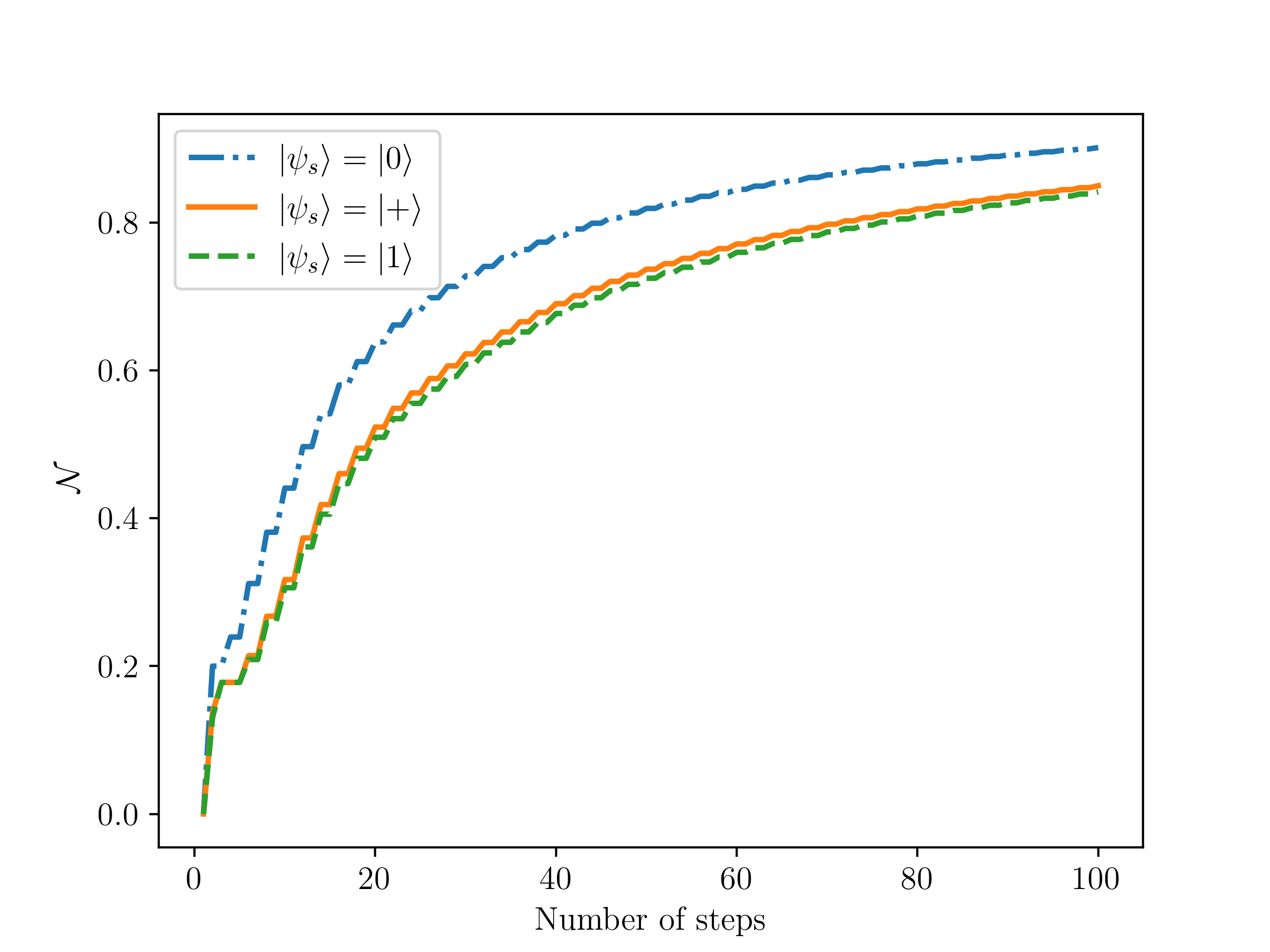}
			\caption{\label{fig:blpvN30_45}}
		\end{subfigure}\\
		\begin{subfigure}{\linewidth}
			\includegraphics[width=9cm]{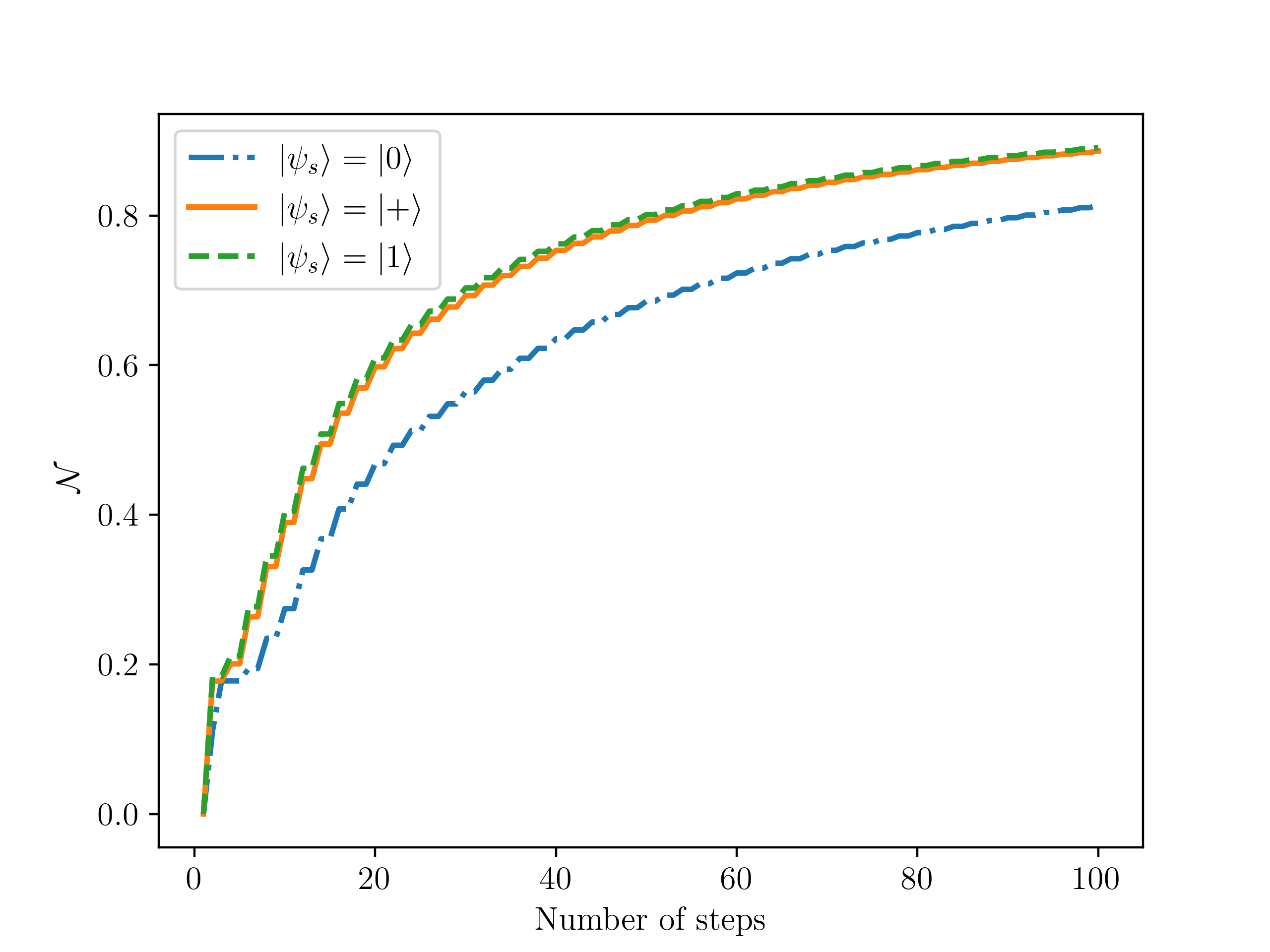}
			\caption{\label{fig:blpvN45_30}}
		\end{subfigure}
		\caption{BLP measure for the cases with definite and equally superposed causal orders of two-period quantum walks. Each walk was executed for $N=100$ steps, with the coin parameters $(\theta_1, \theta_2)$ set to $\left( \frac{\pi}{6}, \frac{\pi}{4} \right)$ and $\left( \frac{\pi}{4}, \frac{\pi}{6} \right)$ in (a) and (b), respectively.}
		\label{fig:parrondo_NM}
	\end{figure}

 A specific temporal order of coin arrangement causes dynamics to be more non-Markovian than the other, in our case the reverse of the former. However, we see that when the dynamics are used in a superposition of temporal order, we observe an interesting aspect peculiar to periodic quantum walks: the Non-Markovianity of the coin dynamics tends to match the trend for one particular causal order. We also see that the degree of non-Markovianity (computed via the BLP measure) under Superposition of causal order is limited by the non-Markovianity of the temporal variant with higher non-Markovian character. 

This effect is illustrated in Figs.~\ref{fig:blpvN30_45} and \ref{fig:blpvN45_30}, where curves corresponding to superposition of causal order are shown in comparison to both forward and reverse causal orders.
\bla
 This is in fact a general trend for quantum walks with higher periods, as can be seen from Fig.~\ref{fig:BLPvPeriods}. \bla
	\begin{figure}
		\includegraphics[width=9cm]{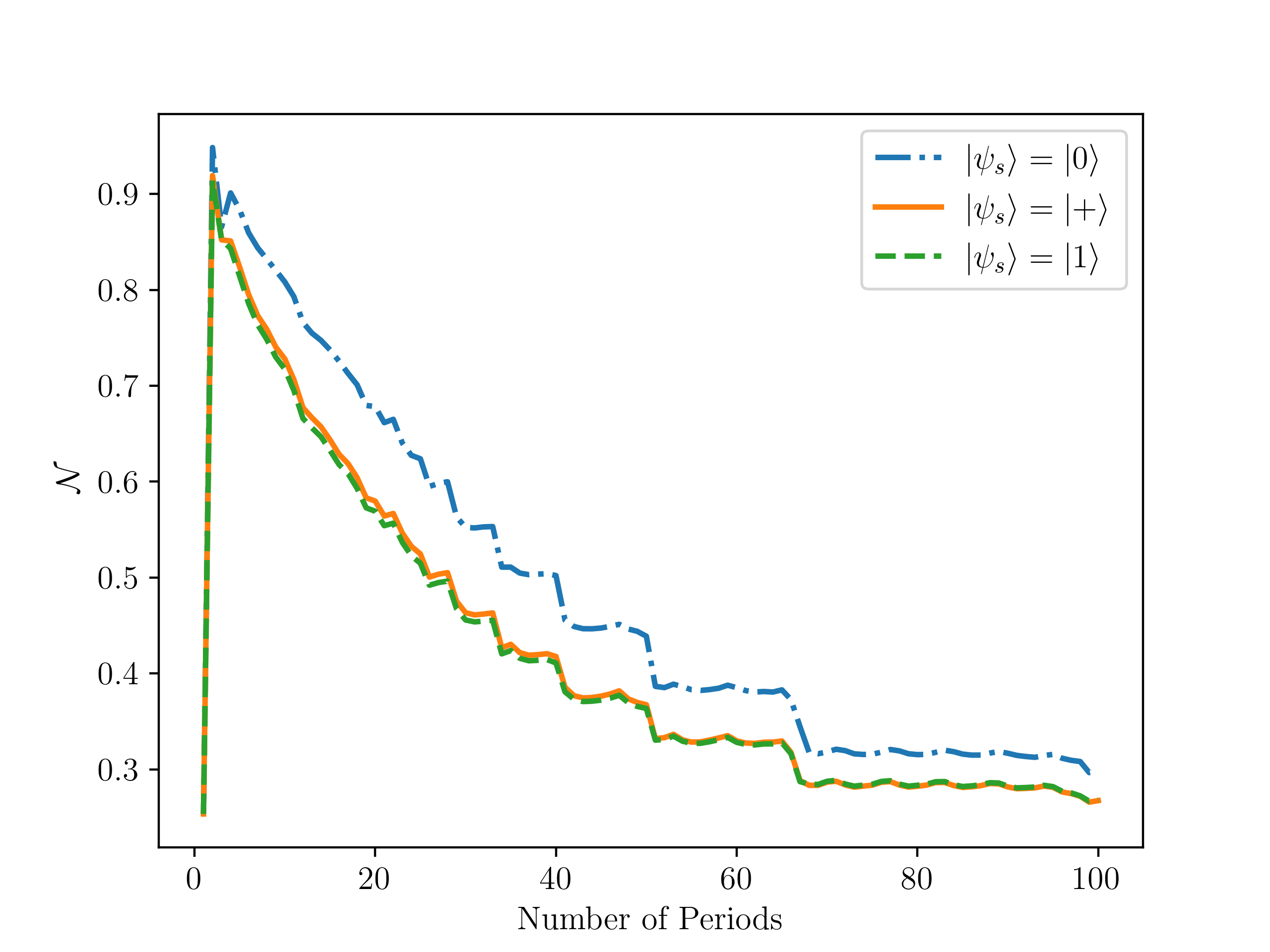}
		\caption{Figure illustrating the normalized BLP measure for a DTQW with different periods, under two definite causal orders, and an equal temporal superposition. The walker has $\theta_1 = \frac{\pi}{6}$, and $\theta_i=\frac{\pi}{4}$ for $i > 1$, and the periodic quantum walk is executed for $N=200$ steps. It is seen that the non-Markovianity of the reduced dynamics of the coin state is bounded by the non-Markovianity of the two causal orders.}
		\label{fig:BLPvPeriods}
	\end{figure}

In conclusion, the present work establishes a framework for the incorporation of Superposition of causal order in quantum walks, delineating specific criteria for its realization. Our findings underscore that only periodic quantum walks or those characterized by at least one disorder manifest Superposition of causal order when subjected to a quantum switch. Through the illustration of a two-period discrete-time quantum walk, we have demonstrated the implications of our criteria, particularly on the reduced dynamics of the coin state. Notably, the inherent causal asymmetry in periodic quantum walks leads to a heightened non-Markovianity in the dynamics of the reduced coin state for one temporal order over the other. Furthermore, we observe a noteworthy alignment between the non-Markovianity induced by indefiniteness in causal order and the dynamics of a particular temporal order of the coin state. This distinctive feature accentuates the impact of Superposition of causal order on the dynamics of quantum walks, shedding light on intriguing facets of quantum systems operating within such paradigms.

\section*{Acknowledgements}
We acknowledge support from the Interdisciplinary Cyber Physical Systems (ICPS) Programme of the Department of Science and Technology, Government of India. Grant No. DST/ICPS/QuST/Theme-1/2019. This work was partially completed while PC and SU were at the Institute of Mathematical Sciences, Chennai. 

    \bla
    \begin{appendix}
    \section{}
    \label{ap:apA}
    \bla
    \begin{Theorem}
        When two completely positive, trace preserving maps are applied in Superposition of causal order, the Kraus rank of the resulting quantum map is at most twice the product of the Kraus ranks of the two maps.
        \label{thm:krausrankproducts}
    \end{Theorem}
    \begin{proof}
        Let $\rho_i$, $\rho_1$ and $\rho_2$ be quantum states such that $\rho_i$ can be transformed into the others via the application of CPTP maps $\Phi_1$ and $\Phi_2$, respectively. In other words,
        \begin{equation}
        \label{eq:maps}
            \begin{split}
                \Phi_1(\rho_i) &= \rho_1, \text{ and,} \\
                \Phi_2(\rho_i) &= \rho_2
            \end{split}
        \end{equation}
        Now, Let $\rho_s$ be a pure state, such that $\rho_s = \ket{\varphi}\bra{\varphi}$, where $\ket{\varphi} = \cos(\theta)\ket{0} + \sin(\theta)\ket{1}$.
        We apply $\Phi_1$ and $\Phi_2$ on $\rho_i$ in temporal superposition by using $\rho_s$ as a quantum switch.
        
        \begin{equation}
		  \mathcal{S}(\Phi_1,\Phi_2)[\rho \otimes \rho_s] = \sum_{i,j} S_{ij} (\rho \otimes \rho_s)S_{ij}^\dagger,
		\label{eq:ICO_channel}
	\end{equation}
	where $\mathcal{S}$ has Kraus operators given as,
	\begin{equation} 
		  S_{ij} = \sum_{i,j} K_i^{(2)}K_j^{(1)} \otimes \ket{0}_s \bra{0} + K_j^{(1)}K_i^{(2)} \otimes \ket{1}_s\bra{1}.
	\end{equation}
        \noindent
        Here $\Phi_j$ has Kraus operators given by the set $\mathcal{K}^{(j)} = \left\{K^{(j)}_i\right\}$, and let $N_{j}$ is the Kraus rank of $\Phi_{j}$ where $j = \{1,2\}$. 

        The switch state $\rho_s = \ket{0}\bra{0}$ or $\rho_s = \ket{1}\bra{1}$ leads to the implementation of $\Phi_1 \Phi_2 [\rho] $ or $\Phi_2 \Phi_1 [\rho]$, respectively. When the switch state is in a superposition $\rho_s = \ket{\psi_s}\bra{\psi_s}$, where $ \ket{\psi_s} = \frac{1}{\sqrt{2}}(\ket{0}+\ket{1})$, the action of the supermap $\mathcal{S}(\Phi_1,\Phi_2)$ shown in Eq.~(\ref{eq:ICO_channel}) leads to non-trivial dynamics of the system $\rho$ by creating a superposition of the two temporal orders of operations.   
        Therefore, we can write,
        \begin{equation}
            \label{eq:krausstate}
            \begin{split}
                \rho_i' = \sum_{m=1}^{N_1}\sum_{n=1}^{N_2} \bigg[ &\cos^2(\theta) K^{(1)}_m K^{(2)}_n \rho_i {K^{(2)}_n}^\dagger {K^{(1)}_m}^\dagger  \\ 
                &+ \sin^2(\theta) K^{(2)}_n K^{(1)}_m \rho_i {K^{(1)}_m}^\dagger {K^{(2)}_n}^\dagger \bigg] 
            \end{split}
        \end{equation}
        Now, consider the set \[ \mathcal{V} = \left(\mathcal{K}^{(1)} \times \mathcal{K}^{(2)} \right) \bigcup \left( \mathcal{K}^{(2)} \times \mathcal{K}^{(1)} \right),\] where \[ \mathcal{K}^{(1)} \times \mathcal{K}^{(2)} = \left\{ \kappa^{(1)}_m \kappa^{(2)}_n \mid \forall \kappa^{(1)}_m \in \mathcal{K}^{(1)},  \kappa^{(2)}_n \in \mathcal{K}^{(2)} \right\}. \] It can be seen that all possible terms in Eq.~\eqref{eq:krausstate} may be written in terms of the elements of $\mathcal{V}$, i.e., 
        \begin{equation}
            \rho_i' = \sum_{m=1}^{\mu}c_m V_m \rho_i V_m^\dagger,
        \end{equation} 
        where $c_m \in \mathbb{C} \, \forall \, m = \{1,2,...,\mu\} $ and $\mu$ is the cardinality of $\mathcal{V}$. 
        Let the operator $\left\lvert X \right\rvert$ denote the cardinality of $X$. Then, we have, 
        \begin{equation}
        \begin{split}
             &\left\lvert \mathcal{K}^{(1)}\right\rvert = N_1 \text{, }  \left\lvert\mathcal{K}^{(2)}\right\rvert = N_2 \\
             \implies &\left\lvert \mathcal{K}^{(1)} \times \mathcal{K}^{(2)} \right\rvert \leq N_1N_2 \\
             \implies &\left\lvert \mathcal{K}^{(2)} \times \mathcal{K}^{(1)} \right\rvert \leq N_2N_1 \\
             \implies &\left\lvert \mathcal{V} \right\rvert \leq 2N_1N_2
        \end{split}
        \end{equation}
        \noindent
        where the inequality in the second and third lines arises due to the consideration that some elements of $\mathcal{K}^{(1)}$ and $\mathcal{K}^{(2)}$ may commute.
    \end{proof}  \\ [1em]

    A straightforward consequence of Theorem~\ref{thm:krausrankproducts} is the fact that two unitary operators when applied in Superposition of causal order can result in an effectively unitary map if and only if they commute.
    \end{appendix}

 \section*{References}


\begin{thebibliography}{42}%
	\makeatletter
	\providecommand \@ifxundefined [1]{%
		\@ifx{#1\undefined}
	}%
	\providecommand \@ifnum [1]{%
		\ifnum #1\expandafter \@firstoftwo
		\else \expandafter \@secondoftwo
		\fi
	}%
	\providecommand \@ifx [1]{%
		\ifx #1\expandafter \@firstoftwo
		\else \expandafter \@secondoftwo
		\fi
	}%
	\providecommand \natexlab [1]{#1}%
	\providecommand \enquote  [1]{``#1''}%
	\providecommand \bibnamefont  [1]{#1}%
	\providecommand \bibfnamefont [1]{#1}%
	\providecommand \citenamefont [1]{#1}%
	\providecommand \href@noop [0]{\@secondoftwo}%
	\providecommand \href [0]{\begingroup \@sanitize@url \@href}%
	\providecommand \@href[1]{\@@startlink{#1}\@@href}%
	\providecommand \@@href[1]{\endgroup#1\@@endlink}%
	\providecommand \@sanitize@url [0]{\catcode `\\12\catcode `\$12\catcode
		`\&12\catcode `\#12\catcode `\^12\catcode `\_12\catcode `\%12\relax}%
	\providecommand \@@startlink[1]{}%
	\providecommand \@@endlink[0]{}%
	\providecommand \url  [0]{\begingroup\@sanitize@url \@url }%
	\providecommand \@url [1]{\endgroup\@href {#1}{\urlprefix }}%
	\providecommand \urlprefix  [0]{URL }%
	\providecommand \Eprint [0]{\href }%
	\providecommand \doibase [0]{https://doi.org/}%
	\providecommand \selectlanguage [0]{\@gobble}%
	\providecommand \bibinfo  [0]{\@secondoftwo}%
	\providecommand \bibfield  [0]{\@secondoftwo}%
	\providecommand \translation [1]{[#1]}%
	\providecommand \BibitemOpen [0]{}%
	\providecommand \bibitemStop [0]{}%
	\providecommand \bibitemNoStop [0]{.\EOS\space}%
	\providecommand \EOS [0]{\spacefactor3000\relax}%
	\providecommand \BibitemShut  [1]{\csname bibitem#1\endcsname}%
	\let\auto@bib@innerbib\@empty
	\bibitem [{\citenamefont {Meyer}(1996)}]{meyer1996quantum}%
	\BibitemOpen
	\bibfield  {author} {\bibinfo {author} {\bibfnamefont {D.~A.}\ \bibnamefont
			{Meyer}},\ }\bibfield  {title} {\bibinfo {title} {From quantum cellular
			automata to quantum lattice gases},\ }\href
	{https://doi.org/10.1007/BF02199356} {\bibfield  {journal} {\bibinfo
			{journal} {Journal of Statistical Physics}\ }\textbf {\bibinfo {volume}
			{85}},\ \bibinfo {pages} {551} (\bibinfo {year} {1996})}\BibitemShut
	{NoStop}%
	\bibitem [{\citenamefont {Venegas-Andraca}(2012)}]{venegas2012quantum}%
	\BibitemOpen
	\bibfield  {author} {\bibinfo {author} {\bibfnamefont {S.~E.}\ \bibnamefont
			{Venegas-Andraca}},\ }\bibfield  {title} {\bibinfo {title} {Quantum walks: a
			comprehensive review},\ }\href {https://doi.org/10.1007/s11128-012-0432-5}
	{\bibfield  {journal} {\bibinfo  {journal} {Quantum Information Processing}\
		}\textbf {\bibinfo {volume} {11}},\ \bibinfo {pages} {1015} (\bibinfo {year}
		{2012})}\BibitemShut {NoStop}%
	\bibitem [{\citenamefont {Chandrashekar}(2012)}]{chandrashekar2012disorder}%
	\BibitemOpen
	\bibfield  {author} {\bibinfo {author} {\bibfnamefont {C.~M.}\ \bibnamefont
			{Chandrashekar}},\ }\bibfield  {title} {\bibinfo {title} {Disorder induced
			localization and enhancement of entanglement in one-and two-dimensional
			quantum walks},\ }\href {https://doi.org/10.48550/arXiv.1212.5984} {\bibfield
		{journal} {\bibinfo  {journal} {arXiv preprint arXiv:1212.5984}\ } (\bibinfo
		{year} {2012})}\BibitemShut {NoStop}%
	\bibitem [{\citenamefont {Mohseni}\ \emph {et~al.}(2008)\citenamefont
		{Mohseni}, \citenamefont {Rebentrost}, \citenamefont {Lloyd},\ and\
		\citenamefont {Aspuru-Guzik}}]{mohseni2008environment}%
	\BibitemOpen
	\bibfield  {author} {\bibinfo {author} {\bibfnamefont {M.}~\bibnamefont
			{Mohseni}}, \bibinfo {author} {\bibfnamefont {P.}~\bibnamefont {Rebentrost}},
		\bibinfo {author} {\bibfnamefont {S.}~\bibnamefont {Lloyd}},\ and\ \bibinfo
		{author} {\bibfnamefont {A.}~\bibnamefont {Aspuru-Guzik}},\ }\bibfield
	{title} {\bibinfo {title} {Environment-assisted quantum walks in
			photosynthetic energy transfer},\ }\href {https://doi.org/10.1063/1.3002335}
	{\bibfield  {journal} {\bibinfo  {journal} {The Journal of chemical physics}\
		}\textbf {\bibinfo {volume} {129}},\ \bibinfo {pages} {11B603} (\bibinfo
		{year} {2008})}\BibitemShut {NoStop}%
	\bibitem [{\citenamefont {Mallick}\ \emph {et~al.}(2017)\citenamefont
		{Mallick}, \citenamefont {Mandal},\ and\ \citenamefont
		{Chandrashekar}}]{mallick2017neutrino}%
	\BibitemOpen
	\bibfield  {author} {\bibinfo {author} {\bibfnamefont {A.}~\bibnamefont
			{Mallick}}, \bibinfo {author} {\bibfnamefont {S.}~\bibnamefont {Mandal}},\
		and\ \bibinfo {author} {\bibfnamefont {C.~M.}\ \bibnamefont
			{Chandrashekar}},\ }\bibfield  {title} {\bibinfo {title} {Neutrino
			oscillations in discrete-time quantum walk framework},\ }\href
	{https://doi.org/10.1140/epjc/s10052-017-4636-9} {\bibfield  {journal}
		{\bibinfo  {journal} {The European Physical Journal C}\ }\textbf {\bibinfo
			{volume} {77}},\ \bibinfo {pages} {1} (\bibinfo {year} {2017})}\BibitemShut
	{NoStop}%
	\bibitem [{\citenamefont {Chawla}\ \emph {et~al.}(2019)\citenamefont {Chawla},
		\citenamefont {Ambarish},\ and\ \citenamefont
		{Chandrashekar}}]{chawla2019quantum}%
	\BibitemOpen
	\bibfield  {author} {\bibinfo {author} {\bibfnamefont {P.}~\bibnamefont
			{Chawla}}, \bibinfo {author} {\bibfnamefont {C.}~\bibnamefont {Ambarish}},\
		and\ \bibinfo {author} {\bibfnamefont {C.~M.}\ \bibnamefont
			{Chandrashekar}},\ }\bibfield  {title} {\bibinfo {title} {Quantum percolation
			in quasicrystals using continuous-time quantum walk},\ }\href
	{https://doi.org/10.1088/2399-6528/ab5ce0} {\bibfield  {journal} {\bibinfo
			{journal} {Journal of Physics Communications}\ }\textbf {\bibinfo {volume}
			{3}},\ \bibinfo {pages} {125004} (\bibinfo {year} {2019})}\BibitemShut
	{NoStop}%
	\bibitem [{\citenamefont {Innocenti}\ \emph {et~al.}(2017)\citenamefont
		{Innocenti}, \citenamefont {Majury}, \citenamefont {Giordani}, \citenamefont
		{Spagnolo}, \citenamefont {Sciarrino}, \citenamefont {Paternostro},\ and\
		\citenamefont {Ferraro}}]{innocenti2017quantum}%
	\BibitemOpen
	\bibfield  {author} {\bibinfo {author} {\bibfnamefont {L.}~\bibnamefont
			{Innocenti}}, \bibinfo {author} {\bibfnamefont {H.}~\bibnamefont {Majury}},
		\bibinfo {author} {\bibfnamefont {T.}~\bibnamefont {Giordani}}, \bibinfo
		{author} {\bibfnamefont {N.}~\bibnamefont {Spagnolo}}, \bibinfo {author}
		{\bibfnamefont {F.}~\bibnamefont {Sciarrino}}, \bibinfo {author}
		{\bibfnamefont {M.}~\bibnamefont {Paternostro}},\ and\ \bibinfo {author}
		{\bibfnamefont {A.}~\bibnamefont {Ferraro}},\ }\bibfield  {title} {\bibinfo
		{title} {Quantum state engineering using one-dimensional discrete-time
			quantum walks},\ }\href {https://doi.org/10.1103/PhysRevA.96.062326}
	{\bibfield  {journal} {\bibinfo  {journal} {Phys. Rev. A}\ }\textbf {\bibinfo
			{volume} {96}},\ \bibinfo {pages} {062326} (\bibinfo {year}
		{2017})}\BibitemShut {NoStop}%
	\bibitem [{\citenamefont {Gerhardt}\ and\ \citenamefont
		{Watrous}(2003)}]{gerhardt2003continuous}%
	\BibitemOpen
	\bibfield  {author} {\bibinfo {author} {\bibfnamefont {H.}~\bibnamefont
			{Gerhardt}}\ and\ \bibinfo {author} {\bibfnamefont {J.}~\bibnamefont
			{Watrous}},\ }\bibfield  {title} {\bibinfo {title} {Continuous-time quantum
			walks on the symmetric group},\ }in\ \href
	{https://doi.org/10.1007/978-3-540-45198-3_25} {\emph {\bibinfo {booktitle}
			{Approximation, Randomization, and Combinatorial Optimization.. Algorithms
				and Techniques}}}\ (\bibinfo  {publisher} {Springer},\ \bibinfo {year}
	{2003})\ pp.\ \bibinfo {pages} {290--301}\BibitemShut {NoStop}%
	\bibitem [{\citenamefont {Godoy}\ and\ \citenamefont
		{Fujita}(1992)}]{godoy1992quantum}%
	\BibitemOpen
	\bibfield  {author} {\bibinfo {author} {\bibfnamefont {S.}~\bibnamefont
			{Godoy}}\ and\ \bibinfo {author} {\bibfnamefont {S.}~\bibnamefont {Fujita}},\
	}\bibfield  {title} {\bibinfo {title} {A quantum random-walk model for
			tunneling diffusion in a 1d lattice. a quantum correction to fick's law},\
	}\href {https://doi.org/10.1063/1.463812} {\bibfield  {journal} {\bibinfo
			{journal} {The Journal of chemical physics}\ }\textbf {\bibinfo {volume}
			{97}},\ \bibinfo {pages} {5148} (\bibinfo {year} {1992})}\BibitemShut
	{NoStop}%
	\bibitem [{\citenamefont {Kitagawa}\ \emph {et~al.}(2010)\citenamefont
		{Kitagawa}, \citenamefont {Rudner}, \citenamefont {Berg},\ and\ \citenamefont
		{Demler}}]{kitagawa2010exploring}%
	\BibitemOpen
	\bibfield  {author} {\bibinfo {author} {\bibfnamefont {T.}~\bibnamefont
			{Kitagawa}}, \bibinfo {author} {\bibfnamefont {M.~S.}\ \bibnamefont
			{Rudner}}, \bibinfo {author} {\bibfnamefont {E.}~\bibnamefont {Berg}},\ and\
		\bibinfo {author} {\bibfnamefont {E.}~\bibnamefont {Demler}},\ }\bibfield
	{title} {\bibinfo {title} {Exploring topological phases with quantum walks},\
	}\href {https://link.aps.org/doi/10.1103/PhysRevA.82.033429} {\bibfield
		{journal} {\bibinfo  {journal} {Physical Review A}\ }\textbf {\bibinfo
			{volume} {82}},\ \bibinfo {pages} {033429} (\bibinfo {year}
		{2010})}\BibitemShut {NoStop}%
	\bibitem [{\citenamefont {Chandrashekar}(2013)}]{chandrashekar2013two}%
	\BibitemOpen
	\bibfield  {author} {\bibinfo {author} {\bibfnamefont {C.~M.}\ \bibnamefont
			{Chandrashekar}},\ }\bibfield  {title} {\bibinfo {title} {Two-component
			dirac-like hamiltonian for generating quantum walk on one-, two-and
			three-dimensional lattices},\ }\href {https://doi.org/10.1038/srep02829}
	{\bibfield  {journal} {\bibinfo  {journal} {Scientific reports}\ }\textbf
		{\bibinfo {volume} {3}},\ \bibinfo {pages} {1} (\bibinfo {year}
		{2013})}\BibitemShut {NoStop}%
	\bibitem [{\citenamefont {Chandrashekar}(2011)}]{chandrashekar2011disordered}%
	\BibitemOpen
	\bibfield  {author} {\bibinfo {author} {\bibfnamefont {C.~M.}\ \bibnamefont
			{Chandrashekar}},\ }\bibfield  {title} {\bibinfo {title}
		{Disordered-quantum-walk-induced localization of a bose-einstein
			condensate},\ }\href {https://doi.org/10.1103/PhysRevA.83.022320} {\bibfield
		{journal} {\bibinfo  {journal} {Physical Review A}\ }\textbf {\bibinfo
			{volume} {83}},\ \bibinfo {pages} {022320} (\bibinfo {year}
		{2011})}\BibitemShut {NoStop}%
	\bibitem [{\citenamefont {Chawla}\ \emph
		{et~al.}(2020{\natexlab{a}})\citenamefont {Chawla}, \citenamefont {Mangal},\
		and\ \citenamefont {Chandrashekar}}]{chawla2020discrete}%
	\BibitemOpen
	\bibfield  {author} {\bibinfo {author} {\bibfnamefont {P.}~\bibnamefont
			{Chawla}}, \bibinfo {author} {\bibfnamefont {R.}~\bibnamefont {Mangal}},\
		and\ \bibinfo {author} {\bibfnamefont {C.~M.}\ \bibnamefont
			{Chandrashekar}},\ }\bibfield  {title} {\bibinfo {title} {Discrete-time
			quantum walk algorithm for ranking nodes on a network},\ }\href
	{https://doi.org/10.1007/s11128-020-02650-4} {\bibfield  {journal} {\bibinfo
			{journal} {Quantum Information Processing}\ }\textbf {\bibinfo {volume}
			{19}},\ \bibinfo {pages} {1} (\bibinfo {year}
		{2020}{\natexlab{a}})}\BibitemShut {NoStop}%
	\bibitem [{\citenamefont {Farhi}\ and\ \citenamefont
		{Gutmann}(1998)}]{farhi1998quantum}%
	\BibitemOpen
	\bibfield  {author} {\bibinfo {author} {\bibfnamefont {E.}~\bibnamefont
			{Farhi}}\ and\ \bibinfo {author} {\bibfnamefont {S.}~\bibnamefont
			{Gutmann}},\ }\bibfield  {title} {\bibinfo {title} {Quantum computation and
			decision trees},\ }\href {https://doi.org/10.1103/PhysRevA.58.915} {\bibfield
		{journal} {\bibinfo  {journal} {Phys. Rev. A}\ }\textbf {\bibinfo {volume}
			{58}},\ \bibinfo {pages} {915} (\bibinfo {year} {1998})}\BibitemShut
	{NoStop}%
	\bibitem [{\citenamefont {Inui}\ \emph {et~al.}(2005)\citenamefont {Inui},
		\citenamefont {Konno},\ and\ \citenamefont {Segawa}}]{konno2005one}%
	\BibitemOpen
	\bibfield  {author} {\bibinfo {author} {\bibfnamefont {N.}~\bibnamefont
			{Inui}}, \bibinfo {author} {\bibfnamefont {N.}~\bibnamefont {Konno}},\ and\
		\bibinfo {author} {\bibfnamefont {E.}~\bibnamefont {Segawa}},\ }\bibfield
	{title} {\bibinfo {title} {One-dimensional three-state quantum walk},\ }\href
	{https://doi.org/10.1103/PhysRevE.72.056112} {\bibfield  {journal} {\bibinfo
			{journal} {Phys. Rev. E}\ }\textbf {\bibinfo {volume} {72}},\ \bibinfo
		{pages} {056112} (\bibinfo {year} {2005})}\BibitemShut {NoStop}%
	\bibitem [{\citenamefont {Yin}\ \emph {et~al.}(2008)\citenamefont {Yin},
		\citenamefont {Katsanos},\ and\ \citenamefont {Evangelou}}]{yin2008quantum}%
	\BibitemOpen
	\bibfield  {author} {\bibinfo {author} {\bibfnamefont {Y.}~\bibnamefont
			{Yin}}, \bibinfo {author} {\bibfnamefont {D.~E.}\ \bibnamefont {Katsanos}},\
		and\ \bibinfo {author} {\bibfnamefont {S.~N.}\ \bibnamefont {Evangelou}},\
	}\bibfield  {title} {\bibinfo {title} {Quantum walks on a random
			environment},\ }\href {https://doi.org/10.1103/PhysRevA.77.022302} {\bibfield
		{journal} {\bibinfo  {journal} {Phys. Rev. A}\ }\textbf {\bibinfo {volume}
			{77}},\ \bibinfo {pages} {022302} (\bibinfo {year} {2008})}\BibitemShut
	{NoStop}%
	\bibitem [{\citenamefont {Douglas}\ and\ \citenamefont
		{Wang}(2008)}]{douglas2008classical}%
	\BibitemOpen
	\bibfield  {author} {\bibinfo {author} {\bibfnamefont {B.~L.}\ \bibnamefont
			{Douglas}}\ and\ \bibinfo {author} {\bibfnamefont {J.~B.}\ \bibnamefont
			{Wang}},\ }\bibfield  {title} {\bibinfo {title} {A classical approach to the
			graph isomorphism problem using quantum walks},\ }\href
	{https://doi.org/10.1088/1751-8113/41/7/075303} {\bibfield  {journal}
		{\bibinfo  {journal} {Journal of Physics A: Mathematical and Theoretical}\
		}\textbf {\bibinfo {volume} {41}},\ \bibinfo {pages} {075303} (\bibinfo
		{year} {2008})}\BibitemShut {NoStop}%
	\bibitem [{\citenamefont {Koll\'ar}\ \emph {et~al.}(2012)\citenamefont
		{Koll\'ar}, \citenamefont {Kiss}, \citenamefont {Novotn\'y},\ and\
		\citenamefont {Jex}}]{kollar2012asymptotic}%
	\BibitemOpen
	\bibfield  {author} {\bibinfo {author} {\bibfnamefont {B.}~\bibnamefont
			{Koll\'ar}}, \bibinfo {author} {\bibfnamefont {T.}~\bibnamefont {Kiss}},
		\bibinfo {author} {\bibfnamefont {J.}~\bibnamefont {Novotn\'y}},\ and\
		\bibinfo {author} {\bibfnamefont {I.}~\bibnamefont {Jex}},\ }\bibfield
	{title} {\bibinfo {title} {Asymptotic dynamics of coined quantum walks on
			percolation graphs},\ }\href {https://doi.org/10.1103/PhysRevLett.108.230505}
	{\bibfield  {journal} {\bibinfo  {journal} {Phys. Rev. Lett.}\ }\textbf
		{\bibinfo {volume} {108}},\ \bibinfo {pages} {230505} (\bibinfo {year}
		{2012})}\BibitemShut {NoStop}%
	\bibitem [{\citenamefont {Childs}(2009)}]{childs2009universal}%
	\BibitemOpen
	\bibfield  {author} {\bibinfo {author} {\bibfnamefont {A.~M.}\ \bibnamefont
			{Childs}},\ }\bibfield  {title} {\bibinfo {title} {Universal computation by
			quantum walk},\ }\href
	{https://link.aps.org/doi/10.1103/PhysRevLett.102.180501} {\bibfield
		{journal} {\bibinfo  {journal} {Physical review letters}\ }\textbf {\bibinfo
			{volume} {102}},\ \bibinfo {pages} {180501} (\bibinfo {year}
		{2009})}\BibitemShut {NoStop}%
	\bibitem [{\citenamefont {Lovett}\ \emph {et~al.}(2010)\citenamefont {Lovett},
		\citenamefont {Cooper}, \citenamefont {Everitt}, \citenamefont {Trevers},\
		and\ \citenamefont {Kendon}}]{lovett2010universal}%
	\BibitemOpen
	\bibfield  {author} {\bibinfo {author} {\bibfnamefont {N.~B.}\ \bibnamefont
			{Lovett}}, \bibinfo {author} {\bibfnamefont {S.}~\bibnamefont {Cooper}},
		\bibinfo {author} {\bibfnamefont {M.}~\bibnamefont {Everitt}}, \bibinfo
		{author} {\bibfnamefont {M.}~\bibnamefont {Trevers}},\ and\ \bibinfo {author}
		{\bibfnamefont {V.}~\bibnamefont {Kendon}},\ }\bibfield  {title} {\bibinfo
		{title} {Universal quantum computation using the discrete-time quantum
			walk},\ }\href {https://doi.org/10.1103/PhysRevA.81.042330} {\bibfield
		{journal} {\bibinfo  {journal} {Physical Review A}\ }\textbf {\bibinfo
			{volume} {81}},\ \bibinfo {pages} {042330} (\bibinfo {year}
		{2010})}\BibitemShut {NoStop}%
	\bibitem [{\citenamefont {Singh}\ \emph {et~al.}(2021)\citenamefont {Singh},
		\citenamefont {Chawla}, \citenamefont {Sarkar},\ and\ \citenamefont
		{Chandrashekar}}]{singh2021universal}%
	\BibitemOpen
	\bibfield  {author} {\bibinfo {author} {\bibfnamefont {S.}~\bibnamefont
			{Singh}}, \bibinfo {author} {\bibfnamefont {P.}~\bibnamefont {Chawla}},
		\bibinfo {author} {\bibfnamefont {A.}~\bibnamefont {Sarkar}},\ and\ \bibinfo
		{author} {\bibfnamefont {C.~M.}\ \bibnamefont {Chandrashekar}},\ }\bibfield
	{title} {\bibinfo {title} {Universal quantum computing using single-particle
			discrete-time quantum walk},\ }\href
	{https://doi.org/10.1038/s41598-021-91033-5} {\bibfield  {journal} {\bibinfo
			{journal} {Scientific Reports}\ }\textbf {\bibinfo {volume} {11}},\ \bibinfo
		{pages} {1} (\bibinfo {year} {2021})}\BibitemShut {NoStop}%
	\bibitem [{\citenamefont {Chawla}\ \emph
		{et~al.}(2020{\natexlab{b}})\citenamefont {Chawla}, \citenamefont {Singh},
		\citenamefont {Agarwal}, \citenamefont {Srinivasan},\ and\ \citenamefont
		{Chandrashekar}}]{chawla2020multi}%
	\BibitemOpen
	\bibfield  {author} {\bibinfo {author} {\bibfnamefont {P.}~\bibnamefont
			{Chawla}}, \bibinfo {author} {\bibfnamefont {S.}~\bibnamefont {Singh}},
		\bibinfo {author} {\bibfnamefont {A.}~\bibnamefont {Agarwal}}, \bibinfo
		{author} {\bibfnamefont {S.}~\bibnamefont {Srinivasan}},\ and\ \bibinfo
		{author} {\bibfnamefont {C.~M.}\ \bibnamefont {Chandrashekar}},\ }\bibfield
	{title} {\bibinfo {title} {Multi-qubit quantum computing using discrete-time
			quantum walks on closed graphs},\ }\href
	{https://doi.org/10.48550/arXiv.2004.05956} {\bibfield  {journal} {\bibinfo
			{journal} {arXiv preprint arXiv:2004.05956}\ } (\bibinfo {year}
		{2020}{\natexlab{b}})}\BibitemShut {NoStop}%
	\bibitem [{\citenamefont {Aharonov}\ \emph {et~al.}(1993)\citenamefont
		{Aharonov}, \citenamefont {Davidovich},\ and\ \citenamefont
		{Zagury}}]{aharonov1993quantum}%
	\BibitemOpen
	\bibfield  {author} {\bibinfo {author} {\bibfnamefont {Y.}~\bibnamefont
			{Aharonov}}, \bibinfo {author} {\bibfnamefont {L.}~\bibnamefont
			{Davidovich}},\ and\ \bibinfo {author} {\bibfnamefont {N.}~\bibnamefont
			{Zagury}},\ }\bibfield  {title} {\bibinfo {title} {Quantum random walks},\
	}\href {https://link.aps.org/doi/10.1103/PhysRevA.48.1687} {\bibfield
		{journal} {\bibinfo  {journal} {Physical Review A}\ }\textbf {\bibinfo
			{volume} {48}},\ \bibinfo {pages} {1687} (\bibinfo {year}
		{1993})}\BibitemShut {NoStop}%
	\bibitem [{\citenamefont {Childs}\ \emph {et~al.}(2003)\citenamefont {Childs},
		\citenamefont {Cleve}, \citenamefont {Deotto}, \citenamefont {Farhi},
		\citenamefont {Gutmann},\ and\ \citenamefont
		{Spielman}}]{childs2003exponential}%
	\BibitemOpen
	\bibfield  {author} {\bibinfo {author} {\bibfnamefont {A.~M.}\ \bibnamefont
			{Childs}}, \bibinfo {author} {\bibfnamefont {R.}~\bibnamefont {Cleve}},
		\bibinfo {author} {\bibfnamefont {E.}~\bibnamefont {Deotto}}, \bibinfo
		{author} {\bibfnamefont {E.}~\bibnamefont {Farhi}}, \bibinfo {author}
		{\bibfnamefont {S.}~\bibnamefont {Gutmann}},\ and\ \bibinfo {author}
		{\bibfnamefont {D.~A.}\ \bibnamefont {Spielman}},\ }\bibfield  {title}
	{\bibinfo {title} {Exponential algorithmic speedup by a quantum walk},\ }in\
	\href {https://doi.org/10.1145/780542.780552} {\emph {\bibinfo {booktitle}
			{Proceedings of the thirty-fifth annual ACM symposium on Theory of
				computing}}}\ (\bibinfo {year} {2003})\ pp.\ \bibinfo {pages}
	{59--68}\BibitemShut {NoStop}%
	\bibitem [{\citenamefont {Brukner}(2014)}]{brukner2014quantum}%
	\BibitemOpen
	\bibfield  {author} {\bibinfo {author} {\bibfnamefont {{\v{C}}.}~\bibnamefont
			{Brukner}},\ }\bibfield  {title} {\bibinfo {title} {Quantum causality},\
	}\href {https://www.nature.com/articles/nphys2930} {\bibfield  {journal}
		{\bibinfo  {journal} {Nature Physics}\ }\textbf {\bibinfo {volume} {10}},\
		\bibinfo {pages} {259} (\bibinfo {year} {2014})}\BibitemShut {NoStop}%
	\bibitem [{\citenamefont {Ebler}\ \emph {et~al.}(2018)\citenamefont {Ebler},
		\citenamefont {Salek},\ and\ \citenamefont {Chiribella}}]{ebler2018enhanced}%
	\BibitemOpen
	\bibfield  {author} {\bibinfo {author} {\bibfnamefont {D.}~\bibnamefont
			{Ebler}}, \bibinfo {author} {\bibfnamefont {S.}~\bibnamefont {Salek}},\ and\
		\bibinfo {author} {\bibfnamefont {G.}~\bibnamefont {Chiribella}},\ }\bibfield
	{title} {\bibinfo {title} {Enhanced communication with the assistance of
			indefinite causal order},\ }\href
	{https://link.aps.org/doi/10.1103/PhysRevLett.120.120502} {\bibfield
		{journal} {\bibinfo  {journal} {Physical review letters}\ }\textbf {\bibinfo
			{volume} {120}},\ \bibinfo {pages} {120502} (\bibinfo {year}
		{2018})}\BibitemShut {NoStop}%
	\bibitem [{\citenamefont {Procopio}\ \emph {et~al.}(2015)\citenamefont
		{Procopio}, \citenamefont {Moqanaki}, \citenamefont {Ara{\'u}jo},
		\citenamefont {Costa}, \citenamefont {Alonso~Calafell}, \citenamefont {Dowd},
		\citenamefont {Hamel}, \citenamefont {Rozema}, \citenamefont {Brukner},\ and\
		\citenamefont {Walther}}]{procopio2015experimental}%
	\BibitemOpen
	\bibfield  {author} {\bibinfo {author} {\bibfnamefont {L.~M.}\ \bibnamefont
			{Procopio}}, \bibinfo {author} {\bibfnamefont {A.}~\bibnamefont {Moqanaki}},
		\bibinfo {author} {\bibfnamefont {M.}~\bibnamefont {Ara{\'u}jo}}, \bibinfo
		{author} {\bibfnamefont {F.}~\bibnamefont {Costa}}, \bibinfo {author}
		{\bibfnamefont {I.}~\bibnamefont {Alonso~Calafell}}, \bibinfo {author}
		{\bibfnamefont {E.~G.}\ \bibnamefont {Dowd}}, \bibinfo {author}
		{\bibfnamefont {D.~R.}\ \bibnamefont {Hamel}}, \bibinfo {author}
		{\bibfnamefont {L.~A.}\ \bibnamefont {Rozema}}, \bibinfo {author}
		{\bibfnamefont {{\v{C}}.}~\bibnamefont {Brukner}},\ and\ \bibinfo {author}
		{\bibfnamefont {P.}~\bibnamefont {Walther}},\ }\bibfield  {title} {\bibinfo
		{title} {Experimental superposition of orders of quantum gates},\ }\href
	{https://doi.org/10.1038/ncomms8913} {\bibfield  {journal} {\bibinfo
			{journal} {Nature communications}\ }\textbf {\bibinfo {volume} {6}},\
		\bibinfo {pages} {1} (\bibinfo {year} {2015})}\BibitemShut {NoStop}%
	\bibitem [{\citenamefont {Rubino}\ \emph {et~al.}(2022)\citenamefont {Rubino},
		\citenamefont {Rozema}, \citenamefont {Massa}, \citenamefont {Ara{\'u}jo},
		\citenamefont {Zych}, \citenamefont {Brukner},\ and\ \citenamefont
		{Walther}}]{rubino2022experimental}%
	\BibitemOpen
	\bibfield  {author} {\bibinfo {author} {\bibfnamefont {G.}~\bibnamefont
			{Rubino}}, \bibinfo {author} {\bibfnamefont {L.~A.}\ \bibnamefont {Rozema}},
		\bibinfo {author} {\bibfnamefont {F.}~\bibnamefont {Massa}}, \bibinfo
		{author} {\bibfnamefont {M.}~\bibnamefont {Ara{\'u}jo}}, \bibinfo {author}
		{\bibfnamefont {M.}~\bibnamefont {Zych}}, \bibinfo {author} {\bibfnamefont
			{{\v{C}}.}~\bibnamefont {Brukner}},\ and\ \bibinfo {author} {\bibfnamefont
			{P.}~\bibnamefont {Walther}},\ }\bibfield  {title} {\bibinfo {title}
		{Experimental entanglement of temporal order},\ }\href
	{https://doi.org/10.22331/q-2022-01-11-621} {\bibfield  {journal} {\bibinfo
			{journal} {Quantum}\ }\textbf {\bibinfo {volume} {6}},\ \bibinfo {pages}
		{621} (\bibinfo {year} {2022})}\BibitemShut {NoStop}%
	\bibitem [{\citenamefont {Taddei}\ \emph {et~al.}(2021)\citenamefont {Taddei},
		\citenamefont {Cari\~ne}, \citenamefont {Mart\'{\i}nez}, \citenamefont
		{Garc\'{\i}a}, \citenamefont {Guerrero}, \citenamefont {Abbott},
		\citenamefont {Ara\'ujo}, \citenamefont {Branciard}, \citenamefont {G\'omez},
		\citenamefont {Walborn}, \citenamefont {Aolita},\ and\ \citenamefont
		{Lima}}]{taddei2021computational}%
	\BibitemOpen
	\bibfield  {author} {\bibinfo {author} {\bibfnamefont {M.~M.}\ \bibnamefont
			{Taddei}}, \bibinfo {author} {\bibfnamefont {J.}~\bibnamefont {Cari\~ne}},
		\bibinfo {author} {\bibfnamefont {D.}~\bibnamefont {Mart\'{\i}nez}}, \bibinfo
		{author} {\bibfnamefont {T.}~\bibnamefont {Garc\'{\i}a}}, \bibinfo {author}
		{\bibfnamefont {N.}~\bibnamefont {Guerrero}}, \bibinfo {author}
		{\bibfnamefont {A.~A.}\ \bibnamefont {Abbott}}, \bibinfo {author}
		{\bibfnamefont {M.}~\bibnamefont {Ara\'ujo}}, \bibinfo {author}
		{\bibfnamefont {C.}~\bibnamefont {Branciard}}, \bibinfo {author}
		{\bibfnamefont {E.~S.}\ \bibnamefont {G\'omez}}, \bibinfo {author}
		{\bibfnamefont {S.~P.}\ \bibnamefont {Walborn}}, \bibinfo {author}
		{\bibfnamefont {L.}~\bibnamefont {Aolita}},\ and\ \bibinfo {author}
		{\bibfnamefont {G.}~\bibnamefont {Lima}},\ }\bibfield  {title} {\bibinfo
		{title} {Computational advantage from the quantum superposition of multiple
			temporal orders of photonic gates},\ }\href
	{https://doi.org/10.1103/PRXQuantum.2.010320} {\bibfield  {journal} {\bibinfo
			{journal} {PRX Quantum}\ }\textbf {\bibinfo {volume} {2}},\ \bibinfo {pages}
		{010320} (\bibinfo {year} {2021})}\BibitemShut {NoStop}%
	\bibitem [{\citenamefont {Chiribella}\ \emph {et~al.}(2021)\citenamefont
		{Chiribella}, \citenamefont {Banik}, \citenamefont {Bhattacharya},
		\citenamefont {Guha}, \citenamefont {Alimuddin}, \citenamefont {Roy},
		\citenamefont {Saha}, \citenamefont {Agrawal},\ and\ \citenamefont
		{Kar}}]{chiribella2021indefinite}%
	\BibitemOpen
	\bibfield  {author} {\bibinfo {author} {\bibfnamefont {G.}~\bibnamefont
			{Chiribella}}, \bibinfo {author} {\bibfnamefont {M.}~\bibnamefont {Banik}},
		\bibinfo {author} {\bibfnamefont {S.~S.}\ \bibnamefont {Bhattacharya}},
		\bibinfo {author} {\bibfnamefont {T.}~\bibnamefont {Guha}}, \bibinfo {author}
		{\bibfnamefont {M.}~\bibnamefont {Alimuddin}}, \bibinfo {author}
		{\bibfnamefont {A.}~\bibnamefont {Roy}}, \bibinfo {author} {\bibfnamefont
			{S.}~\bibnamefont {Saha}}, \bibinfo {author} {\bibfnamefont {S.}~\bibnamefont
			{Agrawal}},\ and\ \bibinfo {author} {\bibfnamefont {G.}~\bibnamefont {Kar}},\
	}\bibfield  {title} {\bibinfo {title} {Indefinite causal order enables
			perfect quantum communication with zero capacity channels},\ }\href
	{https://doi.org/10.1088/1367-2630/abe7a0} {\bibfield  {journal} {\bibinfo
			{journal} {New Journal of Physics}\ }\textbf {\bibinfo {volume} {23}},\
		\bibinfo {pages} {033039} (\bibinfo {year} {2021})}\BibitemShut {NoStop}%
	\bibitem [{\citenamefont {Abbott}\ \emph {et~al.}(2020)\citenamefont {Abbott},
		\citenamefont {Wechs}, \citenamefont {Horsman}, \citenamefont {Mhalla},\ and\
		\citenamefont {Branciard}}]{abbott2020communication}%
	\BibitemOpen
	\bibfield  {author} {\bibinfo {author} {\bibfnamefont {A.~A.}\ \bibnamefont
			{Abbott}}, \bibinfo {author} {\bibfnamefont {J.}~\bibnamefont {Wechs}},
		\bibinfo {author} {\bibfnamefont {D.}~\bibnamefont {Horsman}}, \bibinfo
		{author} {\bibfnamefont {M.}~\bibnamefont {Mhalla}},\ and\ \bibinfo {author}
		{\bibfnamefont {C.}~\bibnamefont {Branciard}},\ }\bibfield  {title} {\bibinfo
		{title} {Communication through coherent control of quantum channels},\ }\href
	{https://doi.org/10.22331/q-2020-09-24-333} {\bibfield  {journal} {\bibinfo
			{journal} {Quantum}\ }\textbf {\bibinfo {volume} {4}},\ \bibinfo {pages}
		{333} (\bibinfo {year} {2020})}\BibitemShut {NoStop}%
	\bibitem [{\citenamefont {Jia}\ and\ \citenamefont
		{Costa}(2019)}]{jia2019causal}%
	\BibitemOpen
	\bibfield  {author} {\bibinfo {author} {\bibfnamefont {D.}~\bibnamefont
			{Jia}}\ and\ \bibinfo {author} {\bibfnamefont {F.}~\bibnamefont {Costa}},\
	}\bibfield  {title} {\bibinfo {title} {Causal order as a resource for quantum
			communication},\ }\href {https://doi.org/10.1103/PhysRevA.100.052319}
	{\bibfield  {journal} {\bibinfo  {journal} {Phys. Rev. A}\ }\textbf {\bibinfo
			{volume} {100}},\ \bibinfo {pages} {052319} (\bibinfo {year}
		{2019})}\BibitemShut {NoStop}%
	\bibitem [{\citenamefont {Milz}\ \emph {et~al.}(2018)\citenamefont {Milz},
		\citenamefont {Pollock}, \citenamefont {Le}, \citenamefont {Chiribella},\
		and\ \citenamefont {Modi}}]{milz2018entanglement}%
	\BibitemOpen
	\bibfield  {author} {\bibinfo {author} {\bibfnamefont {S.}~\bibnamefont
			{Milz}}, \bibinfo {author} {\bibfnamefont {F.~A.}\ \bibnamefont {Pollock}},
		\bibinfo {author} {\bibfnamefont {T.~P.}\ \bibnamefont {Le}}, \bibinfo
		{author} {\bibfnamefont {G.}~\bibnamefont {Chiribella}},\ and\ \bibinfo
		{author} {\bibfnamefont {K.}~\bibnamefont {Modi}},\ }\bibfield  {title}
	{\bibinfo {title} {Entanglement, non-markovianity, and causal
			non-separability},\ }\href {https://doi.org/10.1088/1367-2630/aaafee}
	{\bibfield  {journal} {\bibinfo  {journal} {New Journal of Physics}\ }\textbf
		{\bibinfo {volume} {20}},\ \bibinfo {pages} {033033} (\bibinfo {year}
		{2018})}\BibitemShut {NoStop}%
	\bibitem [{\citenamefont {Utagi}(2021)}]{utagi2021quantum}%
	\BibitemOpen
	\bibfield  {author} {\bibinfo {author} {\bibfnamefont {S.}~\bibnamefont
			{Utagi}},\ }\bibfield  {title} {\bibinfo {title} {Quantum causal correlations
			and non-markovianity of quantum evolution},\ }\href
	{https://doi.org/10.1016/j.physleta.2020.126983} {\bibfield  {journal}
		{\bibinfo  {journal} {Physics Letters A}\ }\textbf {\bibinfo {volume}
			{386}},\ \bibinfo {pages} {126983} (\bibinfo {year} {2021})}\BibitemShut
	{NoStop}%
	\bibitem [{\citenamefont {Giarmatzi}\ and\ \citenamefont
		{Costa}(2021)}]{giarmatzi2021witnessing}%
	\BibitemOpen
	\bibfield  {author} {\bibinfo {author} {\bibfnamefont {C.}~\bibnamefont
			{Giarmatzi}}\ and\ \bibinfo {author} {\bibfnamefont {F.}~\bibnamefont
			{Costa}},\ }\bibfield  {title} {\bibinfo {title} {Witnessing quantum memory
			in non-markovian processes},\ }\href
	{https://doi.org/10.22331/q-2021-04-26-440} {\bibfield  {journal} {\bibinfo
			{journal} {Quantum}\ }\textbf {\bibinfo {volume} {5}},\ \bibinfo {pages}
		{440} (\bibinfo {year} {2021})}\BibitemShut {NoStop}%
	\bibitem [{\citenamefont {Naikoo}\ \emph {et~al.}(2020)\citenamefont {Naikoo},
		\citenamefont {Banerjee},\ and\ \citenamefont
		{Chandrashekar}}]{naikoo2020non}%
	\BibitemOpen
	\bibfield  {author} {\bibinfo {author} {\bibfnamefont {J.}~\bibnamefont
			{Naikoo}}, \bibinfo {author} {\bibfnamefont {S.}~\bibnamefont {Banerjee}},\
		and\ \bibinfo {author} {\bibfnamefont {C.~M.}\ \bibnamefont
			{Chandrashekar}},\ }\bibfield  {title} {\bibinfo {title} {Non-markovian
			channel from the reduced dynamics of a coin in a quantum walk},\ }\href
	{https://doi.org/10.1103/PhysRevA.102.062209} {\bibfield  {journal} {\bibinfo
			{journal} {Physical Review A}\ }\textbf {\bibinfo {volume} {102}},\ \bibinfo
		{pages} {062209} (\bibinfo {year} {2020})}\BibitemShut {NoStop}%
	\bibitem [{\citenamefont {Thompson}\ \emph {et~al.}(2018)\citenamefont
		{Thompson}, \citenamefont {Garner}, \citenamefont {Mahoney}, \citenamefont
		{Crutchfield}, \citenamefont {Vedral},\ and\ \citenamefont
		{Gu}}]{thompson2018causal}%
	\BibitemOpen
	\bibfield  {author} {\bibinfo {author} {\bibfnamefont {J.}~\bibnamefont
			{Thompson}}, \bibinfo {author} {\bibfnamefont {A.~J.~P.}\ \bibnamefont
			{Garner}}, \bibinfo {author} {\bibfnamefont {J.~R.}\ \bibnamefont {Mahoney}},
		\bibinfo {author} {\bibfnamefont {J.~P.}\ \bibnamefont {Crutchfield}},
		\bibinfo {author} {\bibfnamefont {V.}~\bibnamefont {Vedral}},\ and\ \bibinfo
		{author} {\bibfnamefont {M.}~\bibnamefont {Gu}},\ }\bibfield  {title}
	{\bibinfo {title} {Causal asymmetry in a quantum world},\ }\href
	{https://doi.org/10.1103/PhysRevX.8.031013} {\bibfield  {journal} {\bibinfo
			{journal} {Phys. Rev. X}\ }\textbf {\bibinfo {volume} {8}},\ \bibinfo {pages}
		{031013} (\bibinfo {year} {2018})}\BibitemShut {NoStop}%
	\bibitem [{\citenamefont {Kumar}\ \emph {et~al.}(2018)\citenamefont {Kumar},
		\citenamefont {Balu}, \citenamefont {Laflamme},\ and\ \citenamefont
		{Chandrashekar}}]{kumar2018bounds}%
	\BibitemOpen
	\bibfield  {author} {\bibinfo {author} {\bibfnamefont {N.~P.}\ \bibnamefont
			{Kumar}}, \bibinfo {author} {\bibfnamefont {R.}~\bibnamefont {Balu}},
		\bibinfo {author} {\bibfnamefont {R.}~\bibnamefont {Laflamme}},\ and\
		\bibinfo {author} {\bibfnamefont {C.~M.}\ \bibnamefont {Chandrashekar}},\
	}\bibfield  {title} {\bibinfo {title} {Bounds on the dynamics of periodic
			quantum walks and emergence of the gapless and gapped dirac equation},\
	}\href {https://doi.org/10.1103/PhysRevA.97.012116} {\bibfield  {journal}
		{\bibinfo  {journal} {Phys. Rev. A}\ }\textbf {\bibinfo {volume} {97}},\
		\bibinfo {pages} {012116} (\bibinfo {year} {2018})}\BibitemShut {NoStop}%
	\bibitem [{\citenamefont {Rivas}\ \emph {et~al.}(2014)\citenamefont {Rivas},
		\citenamefont {Huelga},\ and\ \citenamefont {Plenio}}]{rivas2014quantum}%
	\BibitemOpen
	\bibfield  {author} {\bibinfo {author} {\bibfnamefont {{\'{A}}.}~\bibnamefont
			{Rivas}}, \bibinfo {author} {\bibfnamefont {S.~F.}\ \bibnamefont {Huelga}},\
		and\ \bibinfo {author} {\bibfnamefont {M.~B.}\ \bibnamefont {Plenio}},\
	}\bibfield  {title} {\bibinfo {title} {Quantum non-markovianity:
			characterization, quantification and detection},\ }\href
	{https://doi.org/10.1088/0034-4885/77/9/094001} {\bibfield  {journal}
		{\bibinfo  {journal} {Reports on Progress in Physics}\ }\textbf {\bibinfo
			{volume} {77}},\ \bibinfo {pages} {094001} (\bibinfo {year}
		{2014})}\BibitemShut {NoStop}%
	\bibitem [{\citenamefont {Li}\ \emph {et~al.}(2018)\citenamefont {Li},
		\citenamefont {Hall},\ and\ \citenamefont {Wiseman}}]{li2018concepts}%
	\BibitemOpen
	\bibfield  {author} {\bibinfo {author} {\bibfnamefont {L.}~\bibnamefont
			{Li}}, \bibinfo {author} {\bibfnamefont {M.~J.}\ \bibnamefont {Hall}},\ and\
		\bibinfo {author} {\bibfnamefont {H.~M.}\ \bibnamefont {Wiseman}},\
	}\bibfield  {title} {\bibinfo {title} {Concepts of quantum non-markovianity:
			A hierarchy},\ }\href {https://doi.org/10.1016/j.physrep.2018.07.001}
	{\bibfield  {journal} {\bibinfo  {journal} {Physics Reports}\ }\textbf
		{\bibinfo {volume} {759}},\ \bibinfo {pages} {1} (\bibinfo {year}
		{2018})}\BibitemShut {NoStop}%
	\bibitem [{\citenamefont {Breuer}\ \emph {et~al.}(2009)\citenamefont {Breuer},
		\citenamefont {Laine},\ and\ \citenamefont {Piilo}}]{breuer2009measure}%
	\BibitemOpen
	\bibfield  {author} {\bibinfo {author} {\bibfnamefont {H.-P.}\ \bibnamefont
			{Breuer}}, \bibinfo {author} {\bibfnamefont {E.-M.}\ \bibnamefont {Laine}},\
		and\ \bibinfo {author} {\bibfnamefont {J.}~\bibnamefont {Piilo}},\ }\bibfield
	{title} {\bibinfo {title} {Measure for the degree of non-markovian behavior
			of quantum processes in open systems},\ }\href
	{https://link.aps.org/doi/10.1103/PhysRevLett.103.210401} {\bibfield
		{journal} {\bibinfo  {journal} {Phys. Rev. Lett.}\ }\textbf {\bibinfo
			{volume} {103}},\ \bibinfo {pages} {210401} (\bibinfo {year}
		{2009})}\BibitemShut {NoStop}%
	\bibitem [{\citenamefont {Filippov}(2021)}]{filippov2021trace}%
	\BibitemOpen
	\bibfield  {author} {\bibinfo {author} {\bibfnamefont {S.~N.}\ \bibnamefont
			{Filippov}},\ }\bibfield  {title} {\bibinfo {title} {Trace decreasing quantum
			dynamical maps: Divisibility and entanglement dynamics},\ }in\ \href@noop {}
	{\emph {\bibinfo {booktitle} {International Conference on Quantum Probability
				\& Related Topics}}}\ (\bibinfo {organization} {Springer},\ \bibinfo {year}
	{2021})\ pp.\ \bibinfo {pages} {121--133}\BibitemShut {NoStop}%
\end{thebibliography}
	\end{document}